\documentclass[english]{article}
\usepackage[T1]{fontenc}
\usepackage[latin9]{inputenc}
\usepackage{graphicx}
\usepackage{esint}

\makeatletter

\newcommand{\lyxdot}{.}

\makeatother

\usepackage{babel}
\begin{document}

\title{Scattering of massive scalars by Schwarzschild black holes}

\author{Derek Lee }

\maketitle
\begin{center}
$\mathit{School}$ $\mathit{of}$ $\mathit{Physics}$ $\mathit{and}$
$\mathit{Astronomy}$ 
\par\end{center}

\begin{center}
$\mathit{University}$ $\mathit{of}$ $\mathit{Minnesota}$
\par\end{center}
\begin{abstract}
The Klein-Gordon equation for the wave function of a single massive
scalar is written in spherical and parabolic coordinates in the presence
of a Schwarzschild background, and some semi-classical techniques
for deriving asymptotic results are considered. In addition to the
well-known logarithmic phase shift due to the tortoise radial coordinate,
it is found that there is a mass-dependent logarithmic phase shift
at infinity. This is found to be necessary to match the Newtonian
cross-section in the non-relativistic limit. Finally, by imposing
suitable boundary conditions near the horizon, the phase shift is
calculated.
\end{abstract}

\section{Introduction}

The problem of a single scalar scattering off a Schwarzschild horizon,
while admittedly academic, is perhaps a useful testing ground for
techniques that might be applied elsewhere. As this case does not
deal with the complication of considering spin, for the most part
ignores possible interactions with surrounding matter, and ignores
any lack of spherical symmetry in the horizon, this situation provides
the opportunity to study how the particle is affected by the presence
of a horizon, without other difficulties obstructing an understanding
of the basic physical processes involved.

\section{General features of the problem}

The equation considered for the purpose of this paper is

\begin{flushleft}
\begin{equation}
\frac{1}{\sqrt{-g}}\partial_{\mu}\left(\sqrt{-g}g^{\mu\nu}\partial_{\nu}\psi\right)=m^{2}\psi\label{eq:1}
\end{equation}
for a scalar with wave function $\psi$, mass $\mathit{m}$, in the
presence of a background metric, written in Schwarzschild spherical
coordinates as
\begin{equation}
g_{\mu\nu}=\left(\begin{array}{cccc}
-\left(1+\Phi\right) & 0 & 0 & 0\\
0 & \left(1+\Phi\right)^{-1} & 0 & 0\\
0 & 0 & r^{2} & 0\\
0 & 0 & 0 & r^{2}\sin^{2}\theta
\end{array}\right)
\end{equation}
where $\Phi=-2GM/r\equiv-R_{S}/r$, for $\mathit{M}$ the mass of
the black hole. Substituting the given metric, one easily sees that
the wave equation becomes
\[
-\left(1+\Phi\right)^{-1}\partial_{t}^{2}\psi+\left(1+\Phi\right)\partial_{r}^{2}\psi+r^{-2}\left[\frac{1}{\sin\theta}\partial_{\theta}\left(\sin\theta\partial_{\theta}\psi\right)+\sin^{-2}\theta\partial_{\phi}^{2}\psi\right]
\]
\begin{equation}
+\frac{2+\Phi}{r}\partial_{r}\psi=m^{2}\psi\label{eq:3}
\end{equation}

\par\end{flushleft}

\begin{figure}
\includegraphics[scale=0.4]{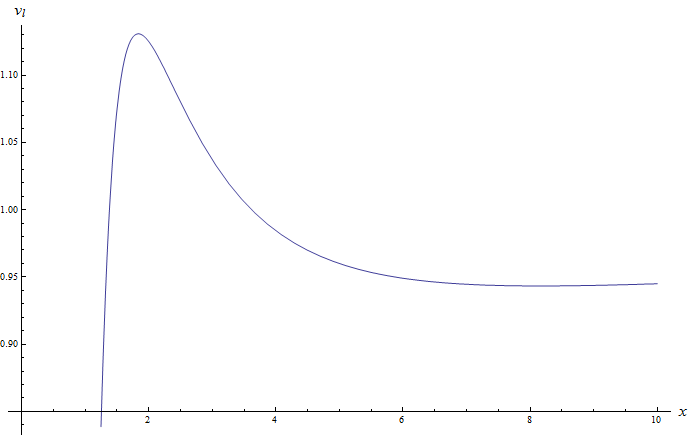}

\caption{\label{fig:potential}$v_{l}/m^{2}$ vs. $x$ for $\alpha=5$}

\end{figure}

Inserting, as usual, the ansatz $\psi=\exp\left(-iEt\right)Y_{lm}R_{l}\left(r\right)$,
this reduces to the radial equation
\begin{equation}
\left(1+\Phi\right)^{2}\frac{d^{2}R_{l}}{dr^{2}}+\frac{\left(2+\Phi\right)\left(1+\Phi\right)}{r}\frac{dR_{l}}{dr}+\left\{ E^{2}-\left[m^{2}+\frac{l\left(l+1\right)}{r^{2}}\right]\left(1+\Phi\right)\right\} R_{l}=0
\end{equation}
Finally, changing variables to the {}``tortoise'' coordinate $r^{*}=\int\frac{dr}{1+\Phi}=r+\ln\left|\frac{r}{R_{S}}-1\right|$
and $\mathit{f_{l}\left(r^{*}\right)=rR_{l}\left(r\right)}$ we find
the {}``Schrodinger'' form,
\begin{equation}
f_{l}^{\prime\prime}+\left(E^{2}-v_{l}\right)f_{l}=0\label{eq:5}
\end{equation}
where
\begin{equation}
v_{l}\equiv m^{2}-\frac{R_{S}m^{2}}{r}+\frac{l\left(l+1\right)}{r^{2}}-\frac{R_{S}\left[l\left(l+1\right)-1\right]}{r^{3}}-\frac{R_{S}^{2}}{r^{4}}\label{eq:6}
\end{equation}
Using the scaled radius $\mathit{x=r/R_{S}}$ and considering the
dimensionless quantity $v_{l}/m^{2}$, it becomes apparent that the
mass only enters in the combination $\left(mR_{S}\right)^{-1}$, that
is, the Compton wavelength of the particle in units of the Schwarzschild
radius. Considering from this point forward only macroscopic black
holes, it is clear that this parameter is extremely small, so that
the angular momentum terms in $\mathit{v_{l}}$ only become important
for very large $\mathit{l}$. In this limit, the distinction between
$l^{2}$, $l\left(l+1\right)$, and $l\left(l+1\right)-1$ can be
ignored. Calling $\left(l/mR_{S}\right)^{2}$$\equiv\alpha$, we finally
have
\begin{equation}
v_{l}/m^{2}=1-\frac{1}{x}+\frac{\alpha}{x^{2}}-\frac{\alpha}{x^{3}}\label{eq:7}
\end{equation}
dropping the (very small) $x^{-4}$ term. (See Figure \ref{fig:potential}.)

Written in this form, the general features of the problem are manifest.
For $r\rightarrow\infty$, the tortoise coordinate approaches the
original radial coordinate (to log accuracy) and the potential approaches
a constant, so that we have $f_{l}\sim\exp\left(\pm ikr\right)$ as
the dominant behavior, where $k^{2}\equiv E^{2}-m^{2}$, of course
as expected. For $r^{*}\rightarrow-\infty$, i.e., $r\rightarrow R_{S}$,
the potential goes to zero and we have $f_{l}\left(r^{*}\right)\exp\left(-iEt\right)\rightarrow\exp\left[-iE\left(t\pm r^{*}\right)\right]$,
independent of the particle's mass.

Looking more closely at the large $\mathit{r}$ limit, some complications
arise. Although the dominant contribution to the phase grows linearly
with distance, the log term in the tortoise radial coordinate introduces
a phase shift which does not tend to a constant at infinity. This
is also very much as expected, as the interaction range is infinite.
We see that the magnitude of this phase shift at large distance is
$kR_{S}\ln\left(r/R_{S}\right)$, as has been noted in the literature
for some time {[}1{]}. However, even after changing the radial coordinate,
we see that there is a Coulomb-type term in the potential, which should
introduce an $\mathit{additional}$ logarithmic phase shift at infinity,
which has not been given sufficient attention thus far. Indeed, introducing
an ansatz for the phase at infinity,
\begin{equation}
\phi\left(r\right)=kr+a\ln\left(r/R_{S}\right)\label{eq:8}
\end{equation}
substituting straight-forwardly into the wave equation and collecting
powers of $1/r$, we find that
\begin{equation}
a=kR_{S}\left(1+\frac{m^{2}}{2k^{2}}\right)\label{eq:9}
\end{equation}
The $\mathit{m}$-dependence does not violate the equivalence principle
as one might at first think {[}2{]}, because the scattering certainly
may (and in fact, does) depend on the incoming particle's velocity
relative to the black hole, which can be expressed in terms of $\mathit{m}$
and $\mathit{k}$.

\section{The Klein-Gordon equation in parabolic coordinates}

\subsection{Solution for forward scattering}

Immediately, the question arises of what boundary conditions to impose
at infinity. For a generic scattering problem, we impose on the wave
function that, at large distance, it resemble a plane wave superimposed
on an outgoing spherical wave. In this case, as with all long-range
potentials, this is too much to ask for, as plane waves become logarithmically
distorted at infinity. As is well known, Coulomb scattering becomes
separable in parabolic coordinates, in which conditions at infinity
are particularly simple to write. Although there is no hope of the
wave equation in this case being completely separable, nevertheless
this is a useful exercise, as scattering at large impact parameter,
i.e., small scattering angle, will only probe the Coulomb tail of
the potential, so the wave equation should be approximately separable
in the forward scattering limit. We will see that this is in fact
the case.

Introducing the coordinates $\eta=r+z,$ $\xi=r-z,$ it is straight-forward
to see that
\begin{equation}
g_{\mu\nu}=\left(\begin{array}{cccc}
-\left(1+\Phi\right) & 0 & 0 & 0\\
0 & \frac{1}{4}\left[\frac{\xi}{\eta}+\left(1+\Phi\right)^{-1}\right] & \frac{1}{4}\left[-1+\left(1+\Phi\right)^{-1}\right] & 0\\
0 & \frac{1}{4}\left[-1+\left(1+\Phi\right)^{-1}\right] & \frac{1}{4}\left[\frac{\eta}{\xi}+\left(1+\Phi\right)^{-1}\right] & 0\\
0 & 0 & 0 & \eta\xi
\end{array}\right)
\end{equation}
where $\Phi=-2R_{S}/\left(\eta+\xi\right)$. Inverting this matrix,
and substituting in (\ref{eq:1}) gives
\[
\frac{4}{\xi+\eta}\left\{ \partial_{\eta}\left[\eta\left(1-\frac{2R_{S}\eta}{\left(\xi+\eta\right)^{2}}\right)\partial_{\eta}\psi\right]+\partial_{\xi}\left[\xi\left(1-\frac{2R_{S}\xi}{\left(\xi+\eta\right)^{2}}\right)\partial_{\xi}\psi\right]\right\} 
\]

\begin{equation}
=\left[m^{2}-\frac{E^{2}}{1+\Phi}\right]\psi\label{eq:11}
\end{equation}
For ordinary Coulomb scattering, the condition imposed on $\psi$
is that it be equal to $\exp\left(ikz\right)f\left(\xi\right)$ for
some function $\mathit{f}$ that is independent of $\eta$ {[}4{]}.
It will be found that this condition is impossible to maintain for
the case at hand. Instead, we can allow $\mathit{f}$ to depend on
both $\xi$ and $\eta$, and examine the limit of large $r=\left(\xi+\eta\right)/2$
and small deflections in the hope that the resulting equations are
then separable.

In this spirit, defining
\begin{equation}
f\left(\eta,\xi\right)=\exp\left[-ik\left(\eta-\xi\right)/2\right]\psi
\end{equation}
and substituting in (\ref{eq:11}) we get, for any coordinate values,\pagebreak{}

\[
\begin{array}{c}
-\frac{R_{S}}{2}\left\{ \frac{E^{2}}{1+\Phi}+k^{2}\left(1-\frac{2\xi\eta}{\left(\xi+\eta\right)^{2}}\right)\right\} f=\end{array}
\]
\[
\eta\left(1+\Phi\frac{\eta}{\xi+\eta}\right)\partial_{\eta}^{2}f+\left[1+ik\eta\left(1+\Phi\frac{\eta}{\eta+\xi}\right)+\Phi\frac{2\xi\eta}{\left(\xi+\eta\right)^{2}}\right]\partial_{\eta}f
\]

\begin{equation}
+\xi\left(1+\Phi\frac{\xi}{\xi+\eta}\right)\partial_{\xi}^{2}f+\left[1-ik\xi\left(1+\Phi\frac{\xi}{\eta+\xi}\right)+\Phi\frac{2\xi\eta}{\left(\xi+\eta\right)^{2}}\right]\partial_{\xi}f\label{eq:13}
\end{equation}
This simplifies for the limits $r/R_{S}\gg1$ and $\theta\rightarrow0$.
The first limit requires us to disregard $\Phi$ terms, the second
to neglect $\frac{\xi\eta}{\left(\xi+\eta\right)^{2}}$, as this is
just $\frac{\sin^{2}\theta}{4}$. In this approximation, (\ref{eq:13})
becomes separable, so that we can finally require $f=f\left(\xi\right)$.
The resulting equation is just
\begin{equation}
\xi f^{\prime\prime}+\left(1-ik\xi\right)f^{\prime}+R_{S}k^{2}\left(1+\frac{m^{2}}{2k^{2}}\right)f=0
\end{equation}
Comparing to the equation for a Coulomb potential {[}4{]}, we see
that this has an identical form, with the replacement $nk\rightarrow-R_{S}k^{2}\left(1+\frac{m^{2}}{2k^{2}}\right)$.
Since the validity of this approximation rests on a large impact parameter,
scattering in this limit should not be significantly affected by boundary
conditions at the horizon. This means that we can deform the potential
close to the horizon to a true Coulomb form without affecting the
forward cross-section; borrowing the usual Rutherford result, we finally
have, for small angles,
\begin{equation}
\frac{d\sigma}{d\Omega}=\frac{R_{S}^{2}}{4\sin^{4}\left(\theta/2\right)}\left(1+\frac{m^{2}}{2k^{2}}\right)^{2}\rightarrow\frac{4R_{S}^{2}}{\theta^{4}}\left(1+\frac{m^{2}}{2k^{2}}\right)^{2}\label{eq:15}
\end{equation}
Comparing with the result in Futterman $\mathit{et}$ $\mathit{al.}$
{[}3{]}, we have precisely the mass dependence anticipated in (\ref{eq:9}).

\subsection{Limiting cases}

It is easy to see that mass dependence of this kind is necessary to
approach the Newtonian scattering limit. Keeping in mind that $n$
is just the potential strength divided by the velocity of the particle
at large separation {[}4{]}, in the Newtonian limit (small velocity,
weak potential, and by extension small angle) the cross-section reduces
to
\begin{equation}
\frac{d\sigma}{d\Omega}=\left(\frac{-mR_{S}/2}{2vk\sin^{2}\left(\theta/2\right)}\right)^{2}=\frac{R_{S}^{2}}{4\sin^{4}\left(\theta/2\right)}\left(\frac{m^{2}}{2k^{2}}\right)^{2}
\end{equation}
matching (\ref{eq:15}) for $m^{2}\gg k^{2}$. In the ultra-relativistic
limit, the cross section loses its momentum dependence, and becomes
a constant $4R_{S}^{2}/\theta^{4}$. The meaning of this becomes clear
when expressed in terms of impact parameter.

Since, by definition,
\begin{equation}
\frac{d\sigma}{d\left(\cos\theta\right)}d\left(\cos\theta\right)=2\pi b\frac{db}{d\theta}d\theta
\end{equation}
we can integrate (\ref{eq:15}) to find $b$. This gives, in the small
angle limit, $\theta=4GM/b$, corresponding exactly to the Einstein
result for light-ray deflection.

\section{The WKB technique}

For larger deflections than those analyzed in the previous section,
we can no longer set $\frac{\xi\eta}{\left(\xi+\eta\right)^{2}}$
to zero, and even for large radii the equation is not separable. Also
the problem arises of setting boundary conditions at the horizon.
The WKB approach is certainly a well-suited alternative for this problem.
The phase in (\ref{eq:8}) certainly changes much more rapidly than
the amplitude or frequency for any macroscopic black hole. However,
the problem of boundary conditions at small $\frac{r}{R_{S}}-1$ becomes
still more pressing, as the phase integral is impossible to even define
without them. 

So far, the physical situation addressed has been unrealistic in a
certain sense. One of the assumptions has been that there is absolutely
no matter in the vicinity of the black hole other than the test particle
in question. Of course, in any realistic case a black hole will be
surrounded by a shell of accreted matter, which will interact with
the scattered particle in an extremely complicated way. We can get
past this obstruction by $\mathit{defining}$ a black hole to be an
object surrounded by an accretion layer so thin that an external particle
is insensitive to the shell's internal dynamics.

Indeed, from (\ref{eq:7}) we see that the potential falls off exponentially
as $r^{*}\rightarrow-\infty$, so that no $l$-dependence is felt
from interaction with potential barriers at large negative $r^{*}$.
Only an overall constant phase shift is left, which has no dynamical
significance. For this restricted class of problems, we can therefore
choose the model of interaction with surrounding matter for convenience.
Choosing an infinite barrier at some fixed radius is especially simple,
and corresponds to choosing a phase of zero at a fixed (small) radius.
This will then be the lower limit of all phase integrals. 

Referring back to (\ref{eq:5}) and (\ref{eq:6}), we see that corrections
to the WKB approximation will be generally of order $\left(kR_{S}\right)^{-1}$
smaller than the dominant terms. By assumption, this is a very small
parameter, and so can be neglected, except very near possible turning
points. Then, keeping in mind the definition $\frac{dr^{*}}{dr}=\frac{1}{1+\Phi}$
and calling $a\sim R_{S}$ the cutoff radius, the phase of the wave-function
is just
\begin{equation}
\phi\left(r\right)=kR_{S}\int_{a/R_{S}}^{x}dx\sqrt{1+\frac{\beta}{x}-\frac{\gamma}{x^{2}}+\frac{\gamma}{x^{3}}}\frac{x}{x-1}
\end{equation}
where $\beta\equiv\frac{m^{2}}{k^{2}}$ and $\gamma\equiv\alpha\beta=\left(l/kR_{S}\right)^{2}$.
By expanding the integrand for large $x$ we have yet another way
of seeing the asymptotic result (\ref{eq:9}).

\begin{figure}
\includegraphics[scale=0.5]{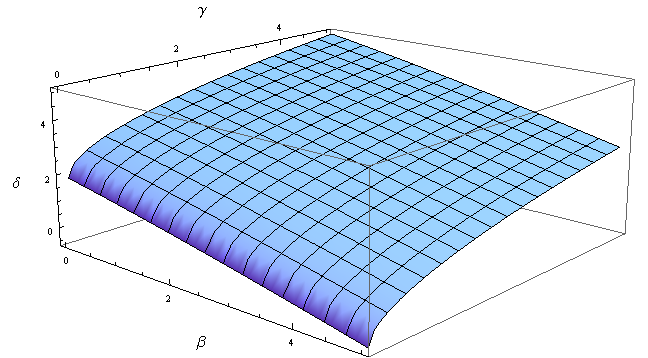}

\caption{\label{fig:2}$\delta$ vs. $\beta$ and $\gamma$ for $x=10$ and
$a/R_{S}=1.05$}

\end{figure}

\begin{figure}
\includegraphics[scale=0.48]{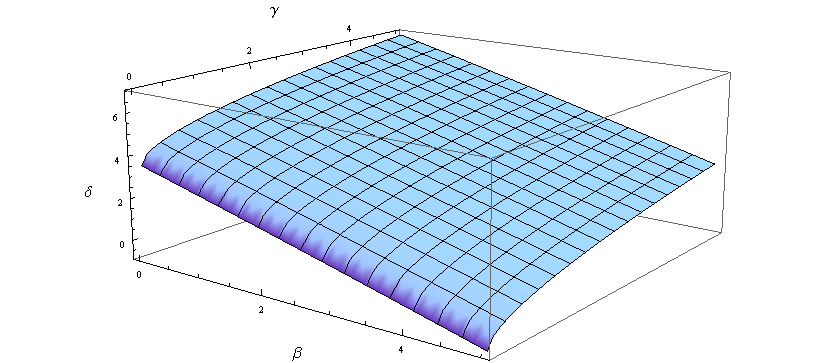}

\caption{\label{fig:3}$\delta$ vs. $\beta$ and $\gamma$ for $x=20$ and
$a/R_{S}=1.01$}

\end{figure}

As we expect the dominant $r$ and $l$ dependence of the phase to
be
\begin{equation}
\varphi\left(r\right)=kR_{S}\left[x-\frac{\pi}{2}\sqrt{\gamma}+\left(1+\frac{\beta}{2}\right)\ln\left(\frac{x}{a/R_{S}}\right)\right]
\end{equation}
{[}4{]}, it is most interesting to look at the behavior of $\delta\equiv\left(\phi-\varphi\right)/kR_{S}$.
For specific values of $a/R_{S}$ and $x$, we can then plot the behavior
of $\delta$ as a function of $\beta$ and $\gamma$. (See figures
\ref{fig:2} and \ref{fig:3}.)

\section{Conclusions}

We see that the case of a massive scalar, as opposed to a massless
one, does in fact modify the results quantitatively, while keeping
all the basic qualitative features. This is not in violation of the
equivalence principle, as the presence of the black hole defines a
preferred reference frame, and the velocity of a particle in this
frame is an important parameter. While it is not possible (or particularly
desirable) to obtain a closed form expression for the cross-section
at arbitrary angle, it is possible to find the forward cross-section
for any value of the remaining parameters, which is the most physically
relevant result for this problem. It has also been found that boundary
conditions at the horizon can be dealt with simply, by treating accreted
matter near the horizon to be an impenetrable barrier. This method
is reasonable, as behavior at infinity is not affected by dynamics
very near the horizon. This method of dealing with the horizon should
be well suited for dealing with more complicated problems, e.g., spinor
or vector scattering, which will be addressed in a future paper.

\section{Acknowledgments}

I would here like to gratefully acknowledge the kind, patient support
and advice of Prof. Tom Walsh, who has provided me with an ideal atmosphere
for work.

\section*{References}

{[}1{]} J. A. H. Futterman, F.A. Handler, and R.A. Matzner, $\mathit{Scattering}$
$\mathit{from}$ $\mathit{Black}$ $\mathit{Holes}$ (Cambridge University
Press, New York, 1988), p. 10.\\
{[}2{]} $\mathit{ibid.}$, p. 79.\\
{[}3{]} $\mathit{ibid}$., p. 97.

\noindent \begin{flushleft}
{[}4{]} L. I. Schiff, $\mathit{Quantum}$ $\mathit{Mechanics}$ (McGraw-Hill,
New York, 1968), 3rd ed., Chap. 5.\linebreak{}

\par\end{flushleft}


\end{document}